\documentclass[aps,twocolumn,showpacs,preprintnumbers,amsmath,amssymb,superscriptaddress]
{revtex4}
\usepackage{graphicx}
\usepackage{dcolumn}
\usepackage{bm}
\usepackage{hyperref}

\def\mk{\mathbf{k}}

\newcommand{\secref}[1]{Sec.~\ref{#1}}
\newcommand{\eqnref}[1]{Eq.~(\ref{#1})}
\newcommand{\figref}[1]{Fig.~\ref{#1}}

\begin{document}

\title
{High Chern number quantum anomalous Hall phases in
graphene ribbons with Haldane orbital coupling}

\author{Tsung-Wei Chen}
\email{twchen@phys.ntu.edu.tw}\affiliation{Department of Physics
and Center for Theoretical Sciences, National Taiwan University,
Taipei 106, Taiwan}

\author{Zhi-Ren Xiao}
\affiliation{Graduate Institute of Applied Physics, National
Chengchi University, Taipei 116, Taiwan}

\author{Dah-Wei Chiou}
\affiliation{Department of Physics and
Center for Theoretical Sciences, National Taiwan University,
Taipei 106, Taiwan}\affiliation{Center for Condensed Matter Sciences, National Taiwan University, Taipei 106, Taiwan}

\author{Guang-Yu Guo}
\email{gyguo@phys.ntu.edu.tw}\affiliation{Department of Physics
and Center for Theoretical Sciences, National Taiwan University,
Taipei 106, Taiwan}\affiliation{Graduate Institute of Applied
Physics, National Chengchi University, Taipei 116, Taiwan}

\date{\today}

\begin{abstract}
We investigate possible phase transitions among the different
quantum anomalous Hall (QAH) phases in a zigzag graphene ribbon under
the influence of the exchange field. 
The effective tight-binding Hamiltonian for graphene is made up
of the hopping term, the Kane-Mele and Rashba spin-orbit
couplings as well as the Haldane orbital term.
We find that the variation of the exchange field results in bulk
gap-closing phenomena and phase transitions occur in the graphene system.
If the Haldane orbital coupling is absent, the phase
transition between the chiral (anti-chiral) edge state $\nu=+2$
($\nu=-2$) and the pseudo-quantum spin Hall state ($\nu=0$) takes place.
Surprisingly, when the Haldane orbital coupling is taken into account,
an intermediate QSH phase with two additional edge modes appears in between
phases $\nu=+2$ and $\nu=-2$. This intermediate phase is therefore either the
hyper-chiral edge state of high Chern number $\nu=+4$ or
anti-hyper-chiral edge state of $\nu=-4$ when the direction
of exchange field is reversed.
We present the band structures, edge state wave functions and
current distributions of the different QAH phases
in the system. We also report the critical exchange field values
for the QAH phase transitions.
\end{abstract}
\pacs{71.70.Ej, 72.25.Dc, 73.43.Nq, 81.05.ue} \maketitle

\section{Introduction}
The anomalous integer quantum Hall effect observed in monolayer
graphenes subjected to an external magnetic
field~\cite{Novoselov05,Zhang05}  has recently attracted considerable attention.
A theoretical investigation~\cite{Gusynin05} showed that plateaus
located at the half-odd-integer position originate from an
additional Landau level at zero energy~\cite{Gusynin06,Peres06},
which is unlike the behavior of the conventional quantum Hall effect
observed in two-dimensional (2D) heterostructure semiconductors.\cite{Klitzing80}
The quantum Hall effect has also
been experimentally observed in AB-stacked bilayer
graphenes~\cite{Novo06}, and this has been studied
theoretically as well.~\cite{Cann06,Nakamura08} Furthermore, much
experimental evidence is available for the existence of AA-stacked
bilayer graphenes.~\cite{Lauffer08} Interestingly, AA-stacked
bilayer graphenes has been shown to exhibit zero transverse
conductivity.~\cite{Fan10}

According to Laughlin's gauge invariance argument, the sample edges
are essential in generating the localized current-carrying states
(edge states)~\cite{Laughlin81,Halperin}. The edge states on
the sample boundary are protected by the bulk band structure
topology which is a manifestation of the Chern number, as
elucidated by Thouless \textit{et al.} (TKNN).~\cite{Thouless} The
TKNN integer $\nu$ (or Chern number) relates the topological class
of the bulk band structure to the number of chiral edge states on
the sample boundary (\emph{bulk-boundary correspondence}) and
hence gives rise to the quantized Hall conductivity $\sigma_{xy}=\nu\,e^2/h$.
The precise quantization of the Hall conductivity arises in the
2D electron system with an integer filling of the Landau
levels. The Chern number corresponding to the number of the chiral
edge currents equals to the number of Landau levels below the Fermi
level. When the system undergoes a phase transition from one chiral
edge state to another, the corresponding Chern number varies
discontinuously from one integer $\nu$ to $\nu\pm 1$.

The Chern number must vanish in a system with time reversal
symmetry (TRS). A TRS breaking mechanism is thus required for
a 2D system to achieve a non-zero Chern number, either with
or without the Landau levels. It has been shown that the chiral
edge state in the quantum Hall phase is related to the
parity anomaly of 2D Dirac fermions.~\cite{Jackiw84, Fradkin86}
Therefore, in a remarkable paper~\cite{Haldane88}, Haldane
constructed a tight-binding Hamiltonian in the 2D honeycomb lattice
with a staggered magnetic field that produces zero average
magnetic flux per unit cell (i.e., no Landau levels) and showed that the
gapped state exhibits the quantum Hall phase with $\nu=\pm1$. In this
sense, the Haldane model is the prototype for the quantum anomalous
Hall (QAH) effect. The relationship between the Chern number and the
winding number of the edge state was investigated in
Ref.~\cite{Hatsugai93}.

On the other hand, when the bulk band gap of a system having a
spin degree of freedom is opened due to the spin-orbit
interaction, the system might be in the quantum spin Hall (QSH)
state where the gapless edge states appearing on the sample
boundary are protected by the TRS.~\cite{KaneRMP10}
The quantization of the spin Hall
conductivity has been predicted in a graphene system with the
Kane-Mele spin-orbit interaction as well as in a semiconductor
superlattice.~\cite{Kane05-1,Kane05-2,Bern06} The
quantization of the spin Hall conductivity, however, may be destroyed by
the parity-breaking perturbations via spin non-conserving term or
disorder. The associated topological invariant classifying the
band structure topology of the time-reversal invariant systems is a
$Z_2$ topological index~\cite{Kane05-1,Kane05-2}. The connection
between the Chern number and the $Z_2$ topological index is
explained in Ref.~\cite{Moore07}. The $Z_2$ topological number
represents the number of the Kramer pairs of the gapless edge modes. An
important result of this classification is that these gapless
edge modes with an odd number of the Kramer pairs in the 2D
systems~\cite{Kane05-1,Kane05-2,Bern06} and an odd number of
surface Dirac cones in the three-dimensional (3D) systems~\cite{Fu07, HZhang09}
are robust to impurity scattering; the other systems are just a
conventional band insulator.

Recently, another topological invariant (i.e., the spin Chern number)
has been proposed by Sheng \textit{et al.}~\cite{Sheng05}, and can be
evaluated by imposing twisted boundary conditions on a finite
sample. In Ref.~\cite{Fukui07}, it has been shown that the spin
Chern number and $Z_2$ topological orders would yield the same
classification by investigating the \emph{bulk gap-closing
phenomena} in the time-reversal invariant systems. The phase
diagram of the 3D QSH systems has been investigated
systematically.~\cite{Murakami08} The topological winding number
related to the spin edge states of graphene with the Kane-Mele
Hamiltonian has also been studied.~\cite{Zhang09} Furthermore, the
bulk-boundary correspondence is generalized to classify
topological defects in insulators and superconductors, where the
gapless boundary excitations are Majorana fermions.~\cite{Kane10}

In this paper, we first model the bulk graphene and also a zigzag
graphene ribbon in the presence of the exchange
field~\cite{Weiss09} by using the Kane-Mele-Rashba Hamiltonian [see
\eqnref{KMR}]. Here the spin degeneracy is lifted by the
TRS breaking term (i.e., the exchange field) and the $z\rightarrow-z$
mirror symmetry is broken by the Rashba term. We calculate
the Chern number of the bulk system as a function of
the exchange field strength. Furthermore, we study
how the edge current in the corresponding graphene ribbon
varies during a phase transition induced by the exchange field.

The quantum anomalous Hall effect in
Hg$_{1-\textrm{y}}$Mn$_\textrm{y}$Te quantum wells~\cite{Liu08}
and tetradymite semiconductors (Bi$_2$Te$_3$, Bi$_2$Se$_3$, and
Sb$_2$Te$_3$)~\cite{Yu10} has been investigated. Graphene with the
Rashba spin-orbit coupling [$\alpha$, see \eqnref{Halpha}] and the
exchange field has also been studied before.~\cite{Qiao10}
However, in Ref. ~\cite{Qiao10}, the Kane-Mele spin-orbit coupling
[$\lambda$, see \eqnref{Hlambda}] was neglected because it was
thought to be weaker than the Rashba spin-orbit interaction. In
this paper, we find that, in the presence of both the Rashba and
Kane-Mele couplings, a phase transition from either a chiral
($\nu=+2$) or anti-chiral ($\nu=-2$) edge state ($\nu=\pm2$) to
the pseudo-QSH state ($\nu=0$) would occur in the graphene ribbon,
because of the change of the Chern number due to the bulk
gap-closing phenomena. This phase transition is different from the
transition between the QSH state and the insulator state when the
exchange field is absent.

We then add the Haldane orbital coupling term which couple the
electron orbital motion to the exchange field~\cite{Haldane88},
to the Kane-Mele-Rashba Hamiltonian for graphene. 
Interestingly, we find that this leads to an
anomalous change in the Chern number pattern. Note that the
Haldane orbital term does not lift the spin degeneracy.
Furthermore, we find that the presence of the Haldane
orbital coupling would give rise to a new intermediate phase between
phases $\nu=+2$ and $\nu=-2$. This intermediate phase has two new
edge modes, and is thus either a hyper-chiral edge state with $\nu=+4$
or an anti-hyper chiral edge state with $\nu=-4$ when the direction
of the exchange field is reversed.

The rest of this paper is organized as follows. In
\secref{sec:Hamiltonian}, we describe the effective tight-binding
Hamiltonian for graphene used in this work. In \secref{sec:Chern number},
we report the energy bands of a graphene ribbon
in the presence of the exchange field. In \secref{sec:KMR system},
we present the phase
transition and the variation of the Chern number with the exchange field in
the Kane-Mele-Rashba system. In particular, we show that the graphene ribbon
undergoes a phase transition from the chiral (or anti-chiral)
state to the pseudo-QSH state. In \secref{sec:HR system}, we
show that a hyper-chiral (or anti-hyper-chiral) state would appear in
between the chiral and anti-chiral states in the Haldane-Rashba
system. The conclusions are given in \secref{sec:con}.


\section{Effective tight-binding Hamiltonian for graphene}\label{sec:Hamiltonian}
\begin{figure}
\begin{center}
\includegraphics[width=8.5cm,height=9cm]{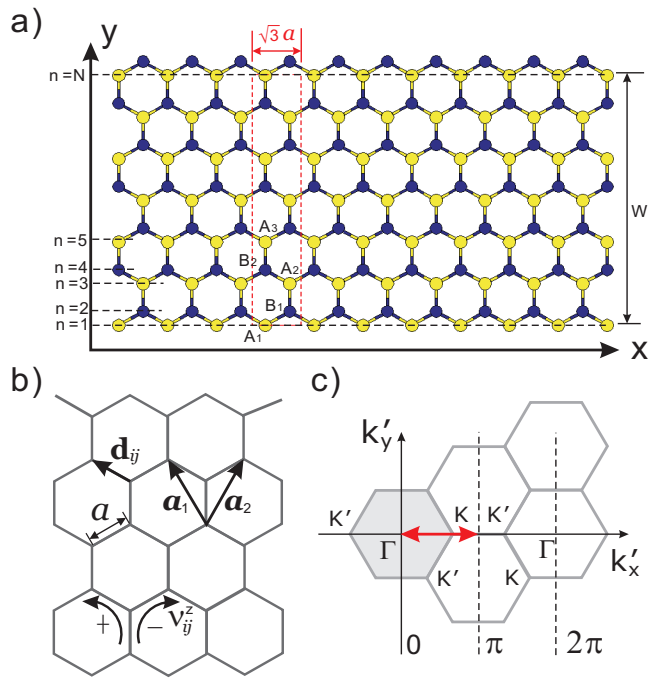}
\end{center}
\caption{(Color online) (a) A segment of a zigzag graphene ribbon with
its unit cell marked by the red dashed lines.
(b) Illustrations of Haldane phase factors $\nu^z_{ij}$, vectors $\mathbf{d}_{ij}$, bulk basis
vectors $(\boldsymbol{a}_1, \boldsymbol{a}_2)$ and bond length $a$.
(c) The first Brillouin zones of bulk graphene (gray region) and the zigzag
graphene ribbon (red double arrow) in the 2D $k$-space.}\label{fig1}
\end{figure}

We consider the effective tight-binding model for graphene given by the
Kane-Mele-Rashba Hamiltonian \cite{Kane05-1, Kane05-2}:
\begin{equation}\label{KMR}
H_{\mathrm{KMR}}=H_t+H_{\lambda}+H_{\alpha},
\end{equation}
with
\begin{subequations}\label{KMGList1}
\begin{align}
&H_t=t\sum_{<i,j>}c_i^{\dag}c_j,\\
&H_{\lambda}=i\lambda\sum_{\ll
i,j\gg}c^{\dag}_is_z\nu_{ij}^zc_{j},\label{Hlambda}\\
&H_{\alpha}=i\alpha\sum_{<
i,j>}c^{\dag}_i(\mathbf{s}\times\mathbf{d}_{ij})_zc_{j}\label{Halpha}.
\end{align}
\end{subequations}
The symbols $<i,j>$ and $\ll i,j\gg$ denote the nearest neighbors
and the next nearest neighbors, respectively. The Hamiltonian $H_t$ is the
tight-binding energy for the nearest-neighbor hopping. The
Kane-Mele Hamiltonian $H_{\lambda}$ describes the intrinsic spin-orbit interaction.
The site-dependent Haldane phase
factor~\cite{Haldane88} $\boldsymbol{\nu}_{ij}$ is defined as
\begin{equation}
\boldsymbol{\nu}_{ij}=\frac{\mathbf{d}_1\times\mathbf{d}_2}{|\mathbf{d}_1\times\mathbf{d}_2|},
\end{equation}
where $\mathbf{d}_i$ denotes the vector from one carbon atom to
one of its nearest neighbors. Two vectors $\mathbf{d}_1$ and
$\mathbf{d}_2$ are required to represent the second
neighbor hopping (see \figref{fig1}). In the two-dimensional case,
the non-zero component $\nu^z_{ij}$ becomes a sign function and we
take the values of $\pm 1$ (i.e., counterclockwise/clockwise).
The extrinsic spin-orbit interaction is described by
the Rashba Hamiltonian $H_{\alpha}$, which can be produced by, e.g., applying
an electric field $\mathbf{E}$ perpendicular to the graphene sheet.
$H_{\alpha}$ is proportional to
$\mathbf{E}\cdot(\mathbf{s}\times\mathbf{d}_{ij})$, where
$\mathbf{E}=E_z\hat{e}_z$ and $\mathbf{d}_{ij}$ denotes the vector
from site $i$ to site $j$ (see \figref{fig1}).

Recent {\it ab initio} density functional calculations showed that
intrinsic ferromagnetism in pure and on-top-Fe-doped graphene monolayers
may exist.~\cite{Qiao10,Son06}
Furthermore, proximity-induced ferromagnetism in graphene was recently reported.~\cite{Tom07}
Therefore, we consider the interaction of the 2D electrons in graphene with
the exchange field produced by the ferromagnetism.~\cite{Weiss09}
The coupling of the orbital motion and also spin of the
electrons on graphene to the exchange field would give rise
to an additional Hamiltonian:
\begin{equation}
H_{\mathrm{ex}}=H_\gamma+H_\beta,
\end{equation}
with
\begin{subequations}\label{KMGList2}
\begin{align}
&H_\gamma=\gamma\sum_{i}c^{\dag}_{i}s_zc_{i},\label{Hgamma}\\
&H_{\beta}=i\tilde{\beta}(\gamma')\sum_{\ll
i,j\gg}c^{\dag}_i\nu^z_{ij}c_{j},\label{Hbeta}
\end{align}
\end{subequations}
where $\gamma$ is the (rescaled) exchange field strength. The
coupling $\gamma$ is proportional to $J_{\mathrm{eff}}\mu'_z$,
where $J_{\mathrm{eff}}$ is the exchange interaction and
$\mu'_z$ is the effective magnetic moment associated with the exchange
field. The magnetic field generated by $\mu'_z$ is denoted as $\gamma'$.
The Hamiltonian $H_\gamma$ describes the response of an electron spin magnetic
moment to the exchange field \textit{\`{a} la} Zeeman effect.

In the meantime, the orbital angular momentum of an electron in graphene
would be coupled to the exchange field because of its associated
orbital magnetic moment.
The Haldane phase factor $\boldsymbol{\nu}_{ij}$ behaves like an
effective orbital angular momentum, and hence gives rise to the interaction
between the electron orbital motion and the magnetic
field $\gamma'$, as described by \eqnref{Hbeta}, where
$\tilde{\beta}(\gamma')$ is a function of $\gamma'$. Spatial
parity symmetry requires $\tilde{\beta}(\gamma')$ to be an odd
function of $\gamma'$. Unlike the interaction between the spin and
exchange field, the energy of the Haldane orbital coupling cannot
be linear in the exchange filed $\gamma'$. Instead, the response
of Haldane orbital motion would be saturated rapidly because the
exchange field $\gamma'$ alters the orbital velocity of electrons and
induces an orbital magnetic moment against it.
Phenomenologically, we can adopt the simple yet sensible approximation:
\begin{equation}
\tilde{\beta}(\gamma')\approx\beta\,\mathrm{sgn}(\gamma),
\end{equation}
where we use $\gamma$ instead of $\gamma'$ for simplicity since
the sign function is independent of the field strength but its
direction. In the present study, the sign of $J_{\mathrm{eff}}$ is fixed,
and hence the sign change of $\gamma$ corresponds to the change in the
direction of $\mu'_z$, which is experimentally possible.
Accordingly, we choose the constant $\beta$ to be negative to have
a diamagnetic response to the magnetic field $\gamma'$.

\section{Chern numbers and edge current chirality}\label{sec:Chern number}
The total Hamiltonian for graphene in the presence of the exchange field
is given by $H=H_\mathrm{KMR}+H_\mathrm{ex}$. For the bulk graphene, the
Hamiltonian $H(\mathbf{k})$ which satisfies the periodicity
$H(\mathbf{k})=H(\mathbf{k}+\mathbf{G})$ ($\mathbf{G}$
stands for a 2D reciprocal-lattice vector), is given by
\begin{equation}\label{bulkH}
H(\mathbf{k})=\left(\begin{array}{cccc}
\lambda'\mathbb{Z}+\gamma&\mathbb{X}+i\mathbb{Y}&0&i\alpha \mathbb{M}_-\\
\mathbb{X}-i\mathbb{Y}&-\lambda'\mathbb{Z}+\gamma&-i\alpha \mathbb{M}_+^{\ast}&0\\
0&i\alpha \mathbb{M}_+&-\lambda''\mathbb{Z}-\gamma&\mathbb{X}+i\mathbb{Y}\\
-i\alpha \mathbb{M}_-^{\ast}&0&\mathbb{X}-i\mathbb{Y}&\lambda''\mathbb{Z}-\gamma\\
\end{array}\right),
\end{equation}
where $\lambda'=\lambda+\beta\,\mathrm{sgn}(\gamma)$ and
$\lambda''=\lambda-\beta\,\mathrm{sgn}(\gamma)$. The state vector
is represented by $\psi^{\dag}= (c^{\dag}_{\mk
A\uparrow},c^{\dag}_{\mk B\uparrow},c^{\dag}_{\mk
A\downarrow},c^{\dag}_{\mk B\downarrow})$, where $A$ and $B$
denote the two different sublattice points in the unit cell,
respectively, and the arrows represent the spin directions. The
matrix elements are given by
$\mathbb{X}=t[1+2\cos(k_x')\cos(3k_y')]$,
$\mathbb{Y}=t[2\cos(k_x')\sin(3k_y')]$,
$\mathbb{Z}=2\sin(2k_x')-4\sin(k_x')\cos(3k_y')$,
$\mathbb{M}_+=[-1+2\cos(k_x'-\frac{\pi}{3})\cos(3k_y')]
+i[2\cos(k_x'-\frac{\pi}{3})\sin(3k_y')]$, and
$\mathbb{M}_-=[-1+2\cos(k_x'+\frac{\pi}{3})\cos(3k_y')]
+i[2\cos(k_x'+\frac{\pi}{3})\sin(3k_y')]$, where the two variables
$k_x^{\prime}$ and $k_{y}^{\prime}$ are defined as
$k_x^{\prime}\equiv\frac{\sqrt{3}}{2}k_xa$ and
$k_y^{\prime}\equiv\frac{k_y}{2}a$, respectively. Note that along
the $k_y=0$ profile, the two points $k_x'=\pm\frac{2\pi}{3}$ are
just the $K$ and $K'$ points in the Brillouin zone of bulk
graphene [see \figref{fig1}(c)], respectively. After the
eigenvalue equation
$H(\mk)|\psi_{n\mk}\rangle=E_{n\mk}|\psi_{n\mk}\rangle$ is solved,
the Berry curvature ($\Omega^{(n)}_{xy}$) of the $n$th band can be
calculated using
\begin{equation}\label{BerryC}
\Omega^{(n)}_{xy}(\mk)=-\sum_{n'(\neq
n)}\frac{2\,\mathrm{Im}\langle\psi_{n\mk}|v_x|\psi_{n'\mk}\rangle\langle\psi_{n'\mk}|v_y|\psi_{n\mk}\rangle}{\left(E_{n'\mk}-E_{n\mk}\right)^2}.
\end{equation}
The Chern number is then obtained by summing the Berry curvatures
$\Omega^{(n)}_{xy}$ for all the occupied states below the Fermi level
for each $\mk$ and subsequently
integrating over the entire first Brillouin zone:
\begin{equation}\label{Chern}
\nu=\frac{1}{2\pi}\sum_{n}\int_{BZ}
dk_xdk_y\Omega^{(n)}_{xy}(\mk).
\end{equation}

The bulk Hamiltonian \eqnref{bulkH} is simplified greatly if we
consider the following simple systems:
\begin{enumerate}
\item Kane-Mele system: $H_\mathrm{KM}=H_t+H_\lambda+H_\gamma$;
\item Rashba system: $H_\mathrm{R}=H_t+H_\alpha+H_\gamma$;
\item Haldane system: $H_\mathrm{H}=H_t+H_\beta+H_\gamma$.
\end{enumerate}

The sign of Chern number indicates the chirality of the edge current. To
verify the occurrence of the edge currents, we compute the energy
band structure for a zigzag graphene ribbon. The unit cell of the zigzag
graphene ribbon is shown in \figref{fig1}(a), where the ribbon
direction is denoted by the $x$ axis and the transverse direction is along
the $y$ direction.
The width of the zigzag ribbon ($W$) is $75\,a$,
where $a$ is the bond length [see \figref{fig1}(b)], i.e., there
are $N+1 = 101$ C atoms in the transverse direction [see \figref{fig1}(a)].
The nearest neighbor hopping integral $t=1$.

\begin{figure}
\begin{center}
\includegraphics[width=8.5cm,height=9.5cm]{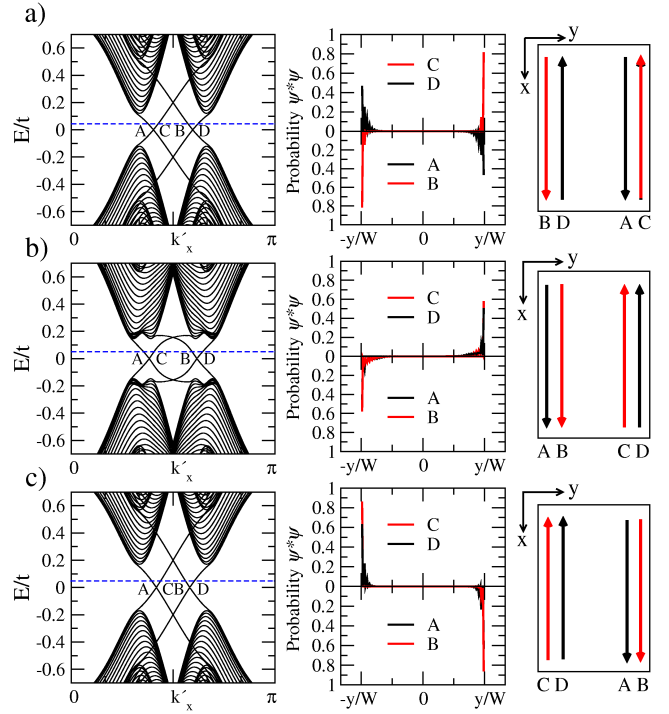}
\end{center}
\caption{(Color online) Calculated energy bands (left panels),
edge state probability (middle panels) and charge
current (right panels) distributions in the zigzag ribbon in the
presence of the exchange field. The Fermi level (the dashed line
in the left panels) $E_F=0.05t$.
(a) Kane-Mele system ($\lambda=0.06t$ and $\gamma=0.2t$). (b)
Rashba system ($\alpha=0.2t$ and $\gamma=0.2t$). (c) Haldane
system ($\beta=-0.07t$ and $\gamma=0.2t$).}\label{fig2}
\end{figure}

Figure 2 shows the ribbon band structure and the edge state probability
distribution in the Kane-Mele
system (a), the Rashba system (b), and the Haldane system (c).
The Fermi level is assumed to be above zero, as indicated by the
dashed horizontal line, and thus, has four intersections
with the conduction bands, denoted as A, B, C, and D, in the left
panels in Fig. 2. This gives rise to four edge currents on the ribbon
edges, as indicated by the A, B, C, and D arrows in the
right panels in Fig. 2. The direction of an edge current, denoted by an arrow,
is given by $I=-|e|v_x$ where the electron group velocity is determined using
$v_x=\partial E_{\mathbf{k}}/\partial k_x$. The $A$ and
$B$ states have the same velocity direction, which is opposite to that of
the $C$ and $D$ states. Hereafter, we use the notation $(I_{L},I_{R})$
to express the charge current distributions on the left-hand side
and right-hand side edges, respectively.
In terms of the bulk-boundary correspondence, for each of the
three systems, the pair A and D would form a single handed loop
(the turning point is at infinity in the $x$ direction), and the pair
B and C would constitute the other loop of the opposite handedness, as
can be seen from the probability distribution shown in the middle
panels in \figref{fig2}.

In the Kane-Mele system $H_\mathrm{KM}$, the current distribution
is $(I_{BD},I_{AC})$, as indicated in the right panel in \figref{fig2}(a).
The two edge states $A$ and $C$ are on the same edge, and so are the $B$ and
$D$ states. As mentioned above, the handedness of the current loop due to
the A and D edge states would produce a Chern number of $-1$ while that
of the pair B and C would give a Chern number of $+1$. Therefore,
the Kane-Mele system is composed of two integer quantum Hall subsystems,
namely, $(\nu=+1)\oplus(\nu=-1)$,~\cite{Kane05-1,XLQ06} and has
$\nu=(+1)+(-1)=0$. Since this state has the same distribution of the edge currents
as that of the quantum spin Hall case with the TRS~\cite{Kane05-1}, except that the
TRS is broken here, we call this state as the pseudo-quantum spin Hall
state.

In the Rashba system $H_\mathrm{R}$, the current distribution is
$(I_{AB},I_{CD})$, as shown in \figref{fig2}(b), which constitute a
paramagnetic response to the exchange field. Both $I_A$ and $I_B$ are
located at the same edge, confirming that the Rashba system has
a Chern number of $+2$, since the two edge current pairs have the same
chirality.~\cite{Qiao10} It is important to note that both the
Chern number and the current distribution $(I_{AB},I_{CD})$ in
the Rashba system is invariant under the transformation
$\alpha\rightarrow-\alpha$. On the other hand, the current distribution becomes
$(I_{CD},I_{AB})$ when the direction of the exchange field is
reversed. Therefore, the Rashba system is equivalent to two integer
quantum Hall subsystems, namely, $(\nu=+1)\oplus(\nu=+1)$ for $\gamma>0$ or
$(\nu=-1)\oplus(\nu=-1)$ for $\gamma<0$.~\cite{Qiao10}

In the Haldane system $H_\mathrm{H}$, the current distribution is
$(I_{CD},I_{AB})$, as shown in \figref{fig2}(c). Both $I_A$ and
$I_B$ are also located at the same edge, but the chirality of the
edge current is opposite to that of the Rashba system, as a result
that the Haldane system with $\beta<0$ exhibits a diamagnetic response to
$\gamma'$. The Chern number of this system is $\nu=-2$. Therefore, the
Haldane system, being diamagnetic, is equivalent to two integer quantum
Hall subsystems, namely, $(\nu=-1)\oplus(\nu=-1)$ for $\gamma>0$
or $(\nu=+1)\oplus(\nu=+1)$ for $\gamma<0$.

In the next two sections, we will consider the following two combinations
of the three simple systems discussed in this section:
\begin{enumerate}
\item Kane-Mele-Rashba system: $H_1=H_{\mathrm{KMR}}+H_\gamma$.
\item Haldane-Rashba system: $H_2=H_t+H_\beta+H_\alpha+H_\gamma$.
\end{enumerate}
We find that, because of the bulk gap-closing phenomena, both
systems will undergo a change of the edge current chirality caused by
varying the exchange field.

\section{Phase transition in the Kane-Mele-Rashba system}\label{sec:KMR system}
In this section, we will neglect the Haldane orbital coupling term of \eqnref{Hbeta}.
We will find that the phase transition is
different from the QSH phase transition in the presence of the
exchange field.
We consider the interplay between $H_{\gamma}$ and
$H_{\mathrm{KMR}}$:
\begin{equation}\label{H1}
\begin{split}
H_1&=H_{\mathrm{KMR}}+H_{\gamma}\\
&=H_t+H_{\alpha}+H_{\lambda}+H_{\gamma}.
\end{split}
\end{equation}
In the presence of both Kane-Mele and Rashba spin-orbit couplings,
the phase transition between the chiral (or anti-chiral) state and
the pseudo-QSH state must occur when the bulk gap-closing
phenomena take place. On the other hand, although the locations of
four currents $(I_{BD},I_{AC})$ become $(I_{AC},I_{BD})$ under the
transformation $\lambda\rightarrow-\lambda$, the phase transition
between (anti-) chiral and pseudo-QSH states still applies because
both $I_{BD}$ and $I_{AC}$ correspond to $\nu=0$.

In order to verify the the phase transition in the
finite system (the graphene ribbon), we use the expectation value of
position $y$ (i.e., $\langle y\rangle$) as a parameter for
specifying the angular momentum of the current in the system,
and define $\langle y\rangle=\langle
y\rangle_A+\langle y\rangle_B$. When the Kane-Mele coupling
$\lambda$ is dominant ($|\gamma|<\gamma^c$, see below), $\langle
y\rangle_A$ and $\langle y\rangle_B$ are on the opposite sides of the
ribbon, and thus $\langle y\rangle/W=0$. When the Rashba coupling
is dominant ($|\gamma|>\gamma^c$, see below), $\langle y\rangle_A$
and $\langle y\rangle_B$ are on the same side of the ribbon. The
quantity $\langle y\rangle/W$ in the Rashba dominant system, however, would
not reach a saturated value $\langle y\rangle/W=\pm1$ owing to the
finite size effect as the perfect edge states is obtained only
when the ribbon width $W$ is infinite. During the phase
transition, the wave function starts to mix with each other in the
central region of the ribbon, and thus, $\langle y\rangle$ is
expected to deviate from either $0$ or $\pm1$.

Let us consider the case of $\lambda=0.06t$, $\alpha=0.05t$ and
$\gamma$ ranging from $-0.5t$ to $0.5t$ in \eqnref{H1}, as an example. For
$\gamma>0$, we find that $\langle y\rangle$ decreases to zero near
some magnitude of $\gamma$ (see below) and does not change
sign, as shown in \figref{fig3}(a). The pattern of $\langle
y\rangle$ in the $\gamma<0$ region is the parity symmetry of that in
$\gamma>0$. We find that the expectation value $\langle y\rangle$
changes sign when the direction of the chiral current is reversed.

\begin{figure}
\begin{center}
\includegraphics[width=8.5cm,height=7.5cm]{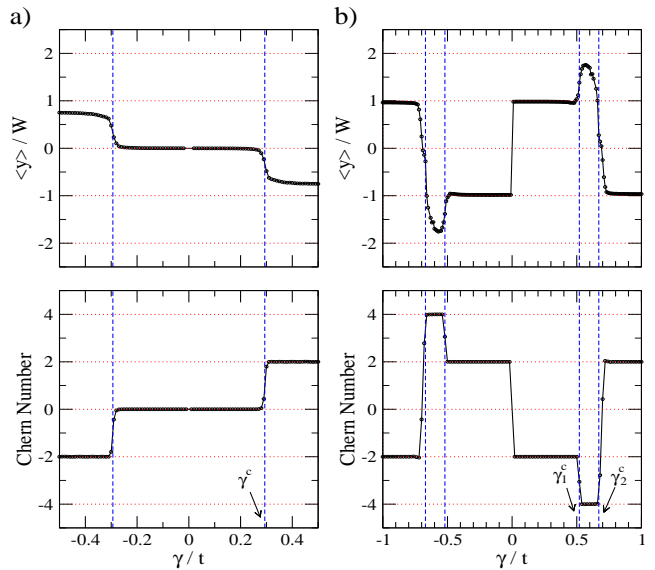}
\end{center}
\caption{(Color online) Expectation value $\langle y\rangle$ and
Chern number as a function of $\gamma/t$. (a) The Kane-Mele-Rashba
system with $\lambda=0.06t$, $\alpha=0.05t$, and $\gamma$ ranging
from $-0.5t$ to $0.5t$. (b) The Haldane-Rashba system with
$\beta=-0.1t$, $\alpha=0.5t$ and $\gamma$ ranging from $-t$ to
$t$.}\label{fig3}
\end{figure}

Based on the bulk-boundary
correspondence~\cite{Thouless,KaneRMP10}, the existence of the phase
transition is supported by evaluating the critical values of the
exchange field. The critical value of the exchange field
($\gamma^c$) for the occurrence of the phase transition is determined
by the bulk gap-closing phenomena, where the bottom of the bulk
conduction band ($E_c$) and the top of the bulk valence band
($E_v$) become degenerate, namely, $E_c-E_v=0$ at $\gamma^c$. It can
be shown that the degenerate point is located at
$k_x'=\pm\frac{2\pi}{3}$. The critical value for the exchange
field is given by
\begin{equation}
\gamma^c=\pm\left(\frac{-\sqrt{3}\alpha^2+12\sqrt{3}\lambda^2}{4\lambda}\right),
\end{equation}
which is obtained for a non-zero Kane-Mele coupling that satisfies
$\gamma/\lambda<3\sqrt{3}$ and $\alpha/\lambda<\sqrt{3}$. The
presence of $\lambda$ causes the critical value for the exchange
field to shift from $\gamma^c=0$ (Rashba system) to a non-zero
value. The magnitude $\gamma^c$ corresponds to the location where
the bulk valence and conduction bands become degenerate and the
Chern number starts to jump from one integer to another.

For a system with given $\alpha$ and $\lambda$, if
$\gamma>\gamma^c$ ($\gamma<\gamma^c$), the system is in the chiral
current state (pseudo-QSH state). When $\gamma=+|\gamma^c|$
(or $\gamma=-|\gamma^c|$), the conduction and valance bands touch
at $k_x'=2\pi/3$ and $k_x'=-2\pi/3$ simultaneously. Because the
$I_A$ current direction is opposite to $I_D$,
the exchange of the locations of $I_A$ and $I_D$ results in
a change of chirality [see \figref{fig3}(a)]. The corresponding
variation of the Chern number is also shown in \figref{fig3}(a).
Therefore, we find that the Chern number jumps from $\nu=0$ to
$\nu=\pm2$. In this case, the critical values $\gamma^c=\pm 0.293t$
are in agreement with the numerical result.

\section{Phase transition in the Haldane-Rashba system}\label{sec:HR system}
In this section, we will neglect the Kane-Mele coupling $\lambda$. We
will show that the presence of the Hamiltonian $H_{\beta}$
creates two new edge modes between the two $\nu = \pm 2$ phases.
Interestingly, these intermediate states are either the hyper-chiral state
($\nu = +4$) or anti-hyper-chiral state ($\nu = -4$).
The Hamiltonian is given by
\begin{equation}\label{H2}
H_2=H_t+H_{\alpha}+H_{\beta}+H_{\gamma}.
\end{equation}
As described in \secref{sec:Chern number}, the Haldane system has
$\nu=-2$ and the Rashba system has $\nu=+2$ when the exchange
field is positive. If the phase transition occurs in this system,
the bulk gap-closing phenomena must take place. In the following,
we show that the Haldane-Rashba system has two
critical values of the exchange field.

For the sake of discussion, we consider the region with
$\gamma>0$, and focus on the region $k_x'>0$ because the
behavior of the corresponding degenerate points in $k_x'<0$ is the
mirror symmetry of that in $k_x'>0$. In the presence of
$\alpha$ only, there are two degenerate points. One is at
$k'_{x1}=2\pi/3$ (i.e., $K$ point), and the other is at
\begin{equation}\label{degen-alpha}
k'_{x2}=\cos^{-1}\left(\frac{2\alpha^2-1}{2\alpha^2+2}\right).
\end{equation}
\eqnref{degen-alpha} shows that the degenerate point
depends on the strength of the coupling and thus varies with the
magnitude of $\alpha$. However, the two degenerate points appear
simultaneously, namely, there is only one critical value for the
exchange field, $\gamma^c=0$, even when the coupling $\lambda$ is
considered [as shown in \figref{fig3}(a)].

Unlike the Kane-Mele coupling $\lambda$, we find that the coupling
$\beta$ will close the bulk gap \emph{twice} at two different
magnitudes of the exchange field. Namely, there are two degenerate
points appearing at two different magnitudes of the exchange field.
Therefore, the Chern number varies discontinuously from $\nu=+2$
to $\nu=-2$ through an intermediate state. One of the degenerate
points is at the $K$ point; the corresponding critical value of
the exchange field is
\begin{equation}
\gamma^c_{1}=-3\sqrt{3}\beta~;~k^{c\prime}_{x1}=2\pi/3.
\end{equation}
However, the second degenerate point for the Hamiltonian
\eqnref{H2} is different from $k'_{x2}$ expressed in
\eqnref{degen-alpha}. The second degenerate point is determined by
the condition $E_v-E_c=0$ and can be expressed as
\begin{equation}
\gamma^c_{2}=-4\sqrt{3}\beta\left(1-F^2\right)~;~k^{c\prime}_{x2}=\cos^{-1}(F),
\end{equation}
where $F$ satisfies the following equation:
\begin{equation}\label{Condition}
-1+2\alpha^2+32\beta^2\left(1-F^2+F^3\right)-2F\left(1+\alpha^2+16\beta^2\right)=0.
\end{equation}
Numerically, \eqnref{Condition} can be solved for a given set of
$\alpha$ and $\beta$. When $\beta=0$, \eqnref{degen-alpha} is the
solution of \eqnref{Condition} and the critical value of the
exchange field is $\gamma^c_{1}=\gamma^c_{2}=0$, which is in
agreement with that in the case of the Rashba system. For the present
case, $\alpha=0.5t$ and $\beta=-0.1t$. The two critical points are
$\gamma^c_{1}=0.5196t$ and $\gamma^c_{2}=0.68995t$. Therefore, the
conduction and valence bands first touch at
$\gamma=\gamma^c_{1}$, and the bulk gap would re-open when
$\gamma^c_{1}<\gamma<\gamma^c_{2}$ (referred to as the
intermediate state). The two bands would touch the second time at
$\gamma=\gamma^c_{2}$. The bulk gap is open when
$\gamma>\gamma^c_{2}$. The calculated Chern number as a
function of $\gamma/t$ is shown in \figref{fig3}(b).

\begin{figure}
\begin{center}
\includegraphics[width=8.5cm,height=9cm]{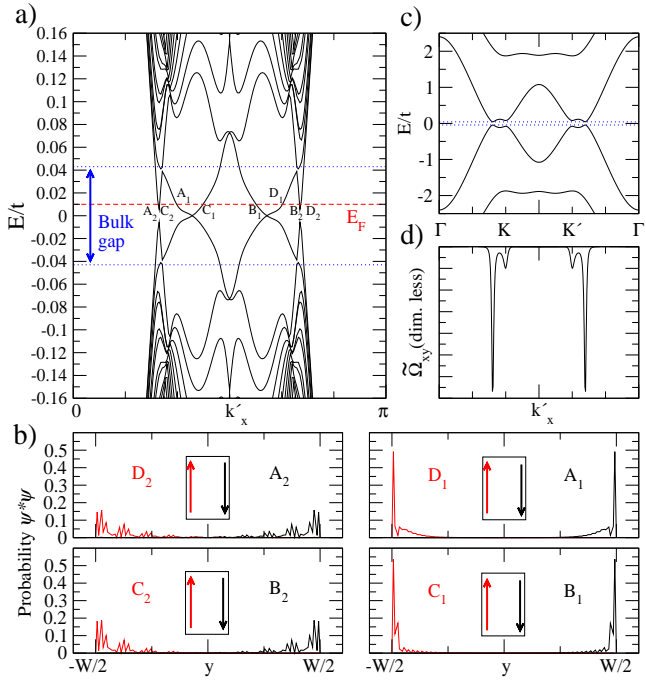}
\end{center}
\caption{(Color online) The band structure (a), the edge state
probability and current distributions (b), bulk energy bands (c)
and Berry curvature (d) of the intermediate state
in $\gamma^c_{1}<\gamma<\gamma^c_{2}$ ($\alpha=0.5t$, $\beta=-0.1t$,
and $\gamma=0.6t$) (see text) in the graphene ribbon.
The Fermi level $E_F$ is at $0.01t$ [the red dashed line in (a)].
In (a), the 8 edge states are marked as $A, B, C, D, A_2, B_2, C_2$,
and $D_2$, respectively.
In (c) and (d), the bulk energy bands and Berry curvature are
plotted along the $k_y=0$ profile.}\label{fig4}
\end{figure}

Surprisingly, the intermediate state between the critical values
$\gamma^c_{1}$ and $\gamma^c_{2}$ shows $\nu=-4$. The band
structure and current distribution of the intermediate state are
shown in \figref{fig4}. We find that in the presence of Haldane
orbital effect, the system establishes two new edge modes: one is
the pair $A_2$ and $D_2$, and the other is the pair $C_2$ and
$B_2$. Furthermore, the current distribution $I_L$ (and $I_R$)
also shows that the four currents in $I_L$ have the same
chirality. Importantly, we find that unlike the quantum Hall
plateau, the Chern number is not restricted in changing from one
integer $\nu$ to the next integer $\nu\pm 1$. Instead, a higher
Chern number can exist in a system with a spin-orbit interaction
if the orbital effect is also taken into account.

The Chern number [see \eqnref{Chern}] can be written as
$\nu=\int_{BZ}dk_x'dk_y'\widetilde{\Omega}_{xy}$. The bulk band
structure and the corresponding Berry curvature
$\widetilde{\Omega}_{xy}$ along the $k_y=0$ profile are shown in
\figref{fig4}(c) and (d), respectively, where we use
$\alpha=0.5t$, $\beta=0.1t$ and $\gamma=0.6t$. It
can be shown that if the linear term
$\beta_1\gamma'$ is taken into account, the Chern number is still $-4$.
This clearly shows that the anomalous Chern number $\nu=-4$ is due
to the second nearest-neighbor hopping in graphene. We
believe that if the third nearest-neighbor hopping is considered
in Haldance orbital effect and spin-orbit interaction, a higher
Chern integer of, e.g., $-6$, may be obtained.

Let us define $\langle y\rangle=\langle
y\rangle_{A_1}+\langle y\rangle_{B_1}+\langle
y\rangle_{A_2}+\langle y\rangle_{B_2}$ for the intermediate state.
The calculated variation of $\langle y\rangle$ with $\gamma$ is shown in \figref{fig3}(b).
Apart from the occurrence of the intermediate state, the expectation
value $\langle y\rangle=\langle y\rangle_A+\langle y\rangle_B$
changes sign as the exchange field is swept through the phase transition,
and this is accompanied by a change of the Chern number
from $\nu=+2$ to $\nu=-2$, as shown in \figref{fig3}(b).
The Chern number obtained from the bulk Hamiltonian represents the number of the perfect
edge states. Note, however, that because of the finite size effect, $\langle
y\rangle/W$ cannot reach the saturated value $\langle
y\rangle/W=\pm 2$ [see \figref{fig3}(b)]. Interestingly, we find that it is not
necessary to reverse the direction of the exchange field in order
to flip the current chirality in this case. Therefore, the
graphene can be brought to either the paramagnetic phase or the
diamagnetic phase by adjusting the magnitude of the exchange field.

Very recently, Tse \textit{et al.}~\cite{Tse11} also proposed
that the Hall conductance can be quantized as $\sigma_{xy}=4e^2/h$,
$albeit$, in bilayer graphene with the Rahsba coupling under the
influence of an external gate voltage. In the present work, in contrast,
we show that in single-layer grapnene, the quantized Hall conductance can
be $\sigma_{xy}=4e^2/h$ when the Haldane orbital effect is
considered.
Furthermore, we find that the change
in the quantized Hall conductance can be achieved by varying the
exchange field instead.\\

\section{Conclusions}\label{sec:con}
In summary, we find that the edge current chirality in a
graphene ribbon can be flipped by varying the exchange field.
The resultant phase
transition of the current chirality is caused by the bulk
gap-closing phenomena; that is, the phase transition is due to the
topological effect of the bulk band structure of graphene.
We show that the paramagnetic response in the Rashba
system can exhibit ether the chiral or anti-chiral state, and thus, the
Hall conductance is quantized as $\sigma_{xy}=\pm2e^2/h$. We find
that the Kane-Mele system has the pseudo-QSH state. For
the Kane-Mele-Rashba system, the transition between the chiral (or
anti-chiral) current state and the pseudo-QSH state can be
achieved by varying the strength of the exchange field.

Unlike the Rashba system, the Haldane system exhibits a
diamagnetic response to the exchange field, and
the quantized Hall conductivity is $\sigma_{xy}=\pm2e^2/h$.
However, the competition between $\alpha$ and $\beta$ leads to a
phase transition between the diamagnetic and paramagnetic responses,
and hence an intermediate phase. Interestingly, this intermediate phase
has two new edge modes and is thus a new quantum anomalous Hall state with
high Chern number $\nu=\pm4$. The corresponding quantized Hall conductance
is $\sigma_{xy}=\pm4e^2/h$ in the graphene ribbon in the absence of Landau levels.

\section*{ACKNOWLEDGMENTS}
T.W.C thanks S. Murakami for valuable
discussions about the construction of the Chern number.
We thank the National Science Council and NCTS of
Taiwan for supports.

\end{document}